\documentclass{PoS_arXiv}
\pdfoutput=1

\usepackage{graphicx} 
\graphicspath{{figs/}} 

\DeclareGraphicsExtensions{.pdf}  

\usepackage{bbm}                  
\usepackage{multirow}             
\usepackage{feynmp}
\usepackage{amsmath}              
\usepackage{amssymb}              
\usepackage{amsthm}               
\usepackage{rotating}             
\usepackage{xcolor}               
\usepackage{cite}

\newcommand{\lsim}
{\;\raisebox{-.3em}{$\stackrel{\displaystyle <}{\sim}$}\;}

\newcommand\Code[1]{\ensuremath{\texttt{#1}}}

\newcommand\TB{t_\beta}

\newcommand\ReDiag{\mathop{%
  \raise .5pt\hbox{[}%
  \widetilde{\mathrm{Re}}%
  \raise .5pt\hbox{]}}}
\newcommand\ReOffDiag{\mathop{%
  \raise .5pt\hbox{$\llbracket$}%
  \widetilde{\mathrm{Re}}%
  \raise .5pt\hbox{$\rrbracket$}}}

\newcommand\MHp{M_{H^\pm}}

\newcommand\Ab{A_b}
\newcommand\At{A_t}

\newcommand\Sn{\tilde\nu}
\newcommand\Sl{\tilde l}

\newcommand\Se{\mathrm{\tilde e}}

\newcommand\Fe{\mathrm{e}}

\newcommand\Fu{\mathrm{u}}

\newcommand\Fd{\mathrm{d}}

\newcommand\cpri{c^\prime}

\newcommand\npri{n^\prime}

\newcommand\spri{s^{\prime}}

\newcommand\ino[1]{\tilde\chi_{#1}}

\newcommand\chapm[1]{\ino{#1}^\pm}
\newcommand\champ[1]{\ino{#1}^\mp}
\newcommand\chap[1]{\ino{#1}^+}
\newcommand\cham[1]{\ino{#1}^-}
\newcommand\cha{\chapm}

\newcommand\neu[1]{\ino{#1}^0}

\newcommand\refta[1]{Tab.~\ref{#1}}

\newcommand\citere[1]{Ref.~\cite{#1}}
\newcommand\citeres[1]{Refs.~\cite{#1}}

\newcommand\ie{i.e.}

\newcommand{\CP}{\cal C\!P}
\newcommand{\cp}{{\CP}}

\newcommand{\onel}{one-loop}
\newcommand{\tev}{\,\, \mathrm{TeV}}
\newcommand{\gev}{\,\, \mathrm{GeV}}

\newcommand{\eecc}{\ensuremath{e^+e^- \to \chapm{c} \champ{\cpri}}}
\newcommand{\eecece}{e^+e^- \to \chap1 \cham1}

\newcommand{\eenn}{\ensuremath{e^+e^- \to \neu{n} \neu{\npri}}}

\newcommand{\eeSlSl}{\ensuremath{e^+e^- \to \Sl_{gs} \Sl_{gs^{\prime}}}}
\newcommand{\eeSeSe}{\ensuremath{e^+e^- \to \Se^{\pm}_{gs} \Se^{\mp}_{gs^{\prime}}}}

\newcommand{\eeSaeSae}{\ensuremath{e^+e^- \to \tilde{\tau}^+_1 \tilde{\tau}^-_1}}

\newcommand{\eeSnSn}{\ensuremath{e^+e^- \to \Sn_{g} \Sn_{g}^*}}

\newcommand\FA{\texttt{FeynArts}}
\newcommand\FC{\texttt{FormCalc}}
\newcommand\LT{\texttt{LoopTools}}

\newcommand\fb{\ensuremath{\mbox{fb}}}

\newcommand{\Sce}{S1}
\newcommand{\Scz}{S2}

\newcommand{\sig}{\sigma}

\newcommand{\phiAeg}{\varphi_{A_{\Fe_g}}}

\def\reffi#1{\mbox{Fig.~\ref{#1}}}

\def\ga{\gamma}

\def\phiAt{\varphi_{\At}}

\def\phiMe{\varphi_{M_1}}

\def\MSL{M_{\tilde L}}
\def\MSE{M_{\tilde E}}

\definecolor{Orange}{named}{orange}
\definecolor{Purple}{named}{purple}
\definecolor{Lightblue}{cmyk}{0.9,0.1,0.1,0.3}
\definecolor{dgelborange}{cmyk}{0.,0.3,0.5, 0.}
\definecolor{Lila}{rgb}{0.5,0.,1}

\allowdisplaybreaks

\title{High-Precision Electroweak SUSY Production Cross Sections at
  \boldmath{$e^+e^-$} Colliders}  

\ShortTitle{EW SUSY Production at $e^+e^-$ Colliders}

\author{\speaker{S.~Heinemeyer}\\ 
Instituto de F\'isica Te\'orica (UAM/CSIC), 
Universidad Aut\'onoma de Madrid, Cantoblanco, 28049, Madrid, Spain\\
        Campus of International Excellence UAM+CSIC, 
Cantoblanco, 28049, Madrid, Spain\\
Instituto de F\'isica de Cantabria (CSIC-UC), 
39005, Santander, Spain\\
        E-mail: \email{Sven.Heinemeyer@cern.ch}}

\author{C.~Schappacher\\
        Institut f\"ur Theoretische Physik,
Karlsruhe Institute of Technology, 
76128, Karlsruhe, Germany (former address)\\
E-mail: \email{schappacher@kabelbw.de}}

\abstract{
For the search for electroweak (EW) particles in the Minimal Supersymmetric 
Standard Model (MSSM) as well as for future precision analyses of these 
particles an accurate knowledge of their production and decay properties
is mandatory. 
We evaluate the cross sections for the chargino, neutralino and
slepton production at $e^+e^-$ colliders in the MSSM with complex
parameters (cMSSM).  
The evaluation is based on a full one-loop calculation of all possible
production channels including soft and hard photon radiation.  
The dependence of the cross sections on the relevant 
cMSSM parameters is analyzed numerically.  We find sizable contributions 
to many production cross sections.  They amount to roughly $15\,\%$ of 
the tree-level results but can go up to $40\,\%$ or higher in extreme 
cases.  Also the dependence on complex parameters of the one-loop 
corrections for many production channels was found non-negligible. 
The full one-loop contributions are thus crucial for physics analyses 
at a future linear $e^+e^-$ collider such as the ILC or CLIC.
}

\FullConference{Loops and Legs in Quantum Field Theory (LL2018)\\
                29 April 2018 - 04 May 2018\\
                St. Goar, Germany}

\begin{document}


\section{Introduction}
\label{sec:intro}

One of the important tasks at the LHC is to search for physics beyond the 
Standard Model (SM), where the Minimal Supersymmetric Standard Model 
(MSSM)~\cite{Ni1984} is one of the leading candidates.
Supersymmetry (SUSY) predicts two scalar partners for all SM fermions as well
as fermionic partners to all SM bosons. 
Contrary to the case of the SM, in the MSSM two Higgs doublets are required.
This results in five physical Higgs bosons instead of the single Higgs
boson in the SM.  These are the light and heavy $\cp$-even Higgs bosons, 
$h$ and $H$, the $\cp$-odd Higgs boson, $A$, and the charged Higgs bosons,
$H^\pm$.
The neutral SUSY partners of the (neutral) Higgs and electroweak gauge
bosons are the four neutralinos, $\neu{1,2,3,4}$.  The corresponding
charged SUSY partners are the charginos, $\cha{1,2}$.
The SUSY partner of the charged leptons are the 
$\tilde e_s, \tilde\mu_s, \tilde\tau_s$ ($s = 1,2$), the ones of the
neutrinos are the $\Sn_e, \Sn_\mu, \Sn_\tau$.

If SUSY is realized in nature and the scalar quarks and/or the gluino
are in the kinematic reach of the (HL-)LHC, it is expected that these
strongly interacting particles are eventually produced and studied.
On the other hand, SUSY particles that interact only via the electroweak
force, \ie, the charginos, neutralinos, and scalar leptons, have a much 
smaller production cross section at the LHC.  Correspondingly, the LHC
discovery potential as well as the current experimental bounds are
substantially weaker~\cite{ATLAS-SUSY,CMS-SUSY}.
At a (future) $e^+e^-$ collider charginos, neutralinos and sleptons,
depending on their masses  
and the available center-of-mass energy, could be produced and
analyzed in detail~\cite{MC15,MC-LCWS17}.  
Corresponding studies can be found for the ILC 
in \citeres{ILC-TDR,LCreport} and for CLIC 
in \citeres{CLIC1,LCreport}. 
(Results on the combination of LHC and ILC results can be found in 
\citere{lhcilc1}.)  Such precision studies will be 
crucial to determine their nature and the underlying SUSY parameters.

In order to yield a sufficient accuracy, one-loop corrections to the 
various production and decay modes have to be considered.
Full one-loop calculations in the cMSSM of (heavy) scalar tau decays was
evaluated in \citere{Stau2decay}, where the calculation can easily be taken
over to other slepton decays. Similarly, full one-loop calculations 
for various chargino/neutralino decays in the cMSSM have been presented in
\citere{LHCxC}.
Sleptons can also be produced in SUSY cascade decays, where full one-loop
evaluations in the cMSSM exist for the corresponding decays of Higgs
bosons~\cite{HiggsDecaySferm}. 
Similarly, the one-loop corrections for chargino/neutralino production
from the decay of Higgs bosons (at the LHC or ILC/CLIC) can be found
in \citere{HiggsDecayIno}.
Here we review the predictions for chargino, neutralino and slepton
production at $e^+e^-$ colliders~\cite{eeIno,eeSlep} (see also 
\citere{eeLCWS17}), \ie\ the channels
(with $\Se_{gs} = \{\tilde e_s, \tilde\mu_s, \tilde\tau_s\}$, 
$\Sn_g = \{\Sn_e, \Sn_\mu, \Sn_\tau\}$, generation index $g$ and slepton 
index $s$)
\begin{alignat}{4}
\label{eq:eecc}
&\sig(\eecc) &~~c,\cpri = 1,2\,, \qquad 
&\sig(\eenn) &~~n,\npri = 1,2,3,4\,, \\
&\sig(\eeSeSe) &s,\spri = 1,2\,, \qquad 
\label{eq:eeSnSn}
&\sig(\eeSnSn) &g = 1,2,3\,.
\end{alignat}


\section{Calculation of diagrams}
\label{sec:calc}

In this section we review some details regarding the renormalization
procedure and the calculation of the tree-level and higher-order 
corrections to the production of charginos, neutralinos and sleptons
in $e^+e^-$ collisions.  
The diagrams and corresponding amplitudes have been obtained with \FA\ 
(version 3.9) \cite{feynarts1}, using our MSSM model 
file (including the MSSM counterterms) of \citere{MSSMCT}. 
The further evaluation has been performed with \FC\ (version 9.5) and 
\LT\ (version 2.14)~\cite{formcalc1}.

The cross sections (\ref{eq:eecc}) - (\ref{eq:eeSnSn}) are calculated 
at the one-loop level, including soft, hard and collinear QED radiation.
This requires the simultaneous renormalization of the 
gauge-boson sector, the fermion/sfermion sector as well as the
chargino/neutralino sector of the  cMSSM, based on \citeres{MSSMCT,mhcMSSMlong}.
All the relevant details can
be found in \citeres{eeIno,eeSlep}. 
The renormalization scheme employed is the same one as for the decay 
of sleptons~\cite{Stau2decay} or
charginos/neutralinos~\cite{LHCxC}. 
Consequently, the predictions for the production and decay can be 
used together in a consistent manner.
More details and the application to Higgs-boson and SUSY  
particle decays can be found in 
\citeres{HiggsDecaySferm,HiggsDecayIno,MSSMCT,SbotRen1,Stop2decay,%
Gluinodecay,Stau2decay,LHCxC}.  Similarly, the 
application to Higgs-boson production cross sections 
at $e^+e^-$ colliders are given in \citeres{HiggsProd,HpProd}.

\begin{figure}
\begin{center}
\framebox[14cm]{\includegraphics[width=0.24\textwidth]{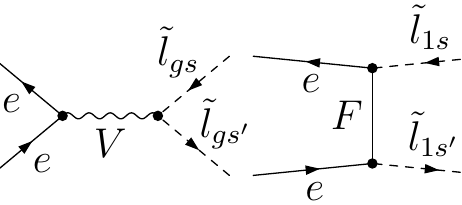}}
\framebox[14cm]{\includegraphics[width=0.75\textwidth]{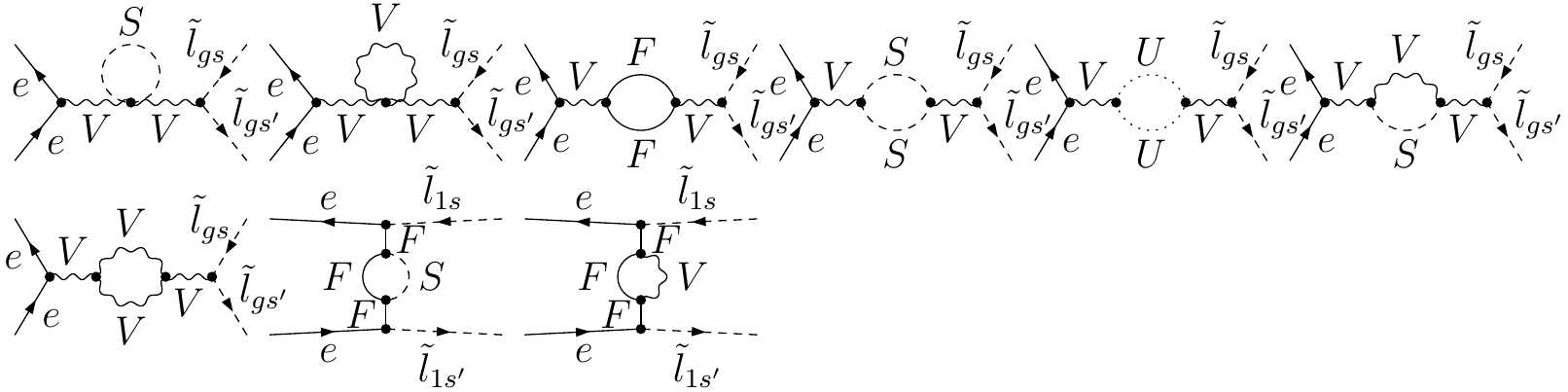}}
\framebox[14cm]{\includegraphics[width=0.75\textwidth]{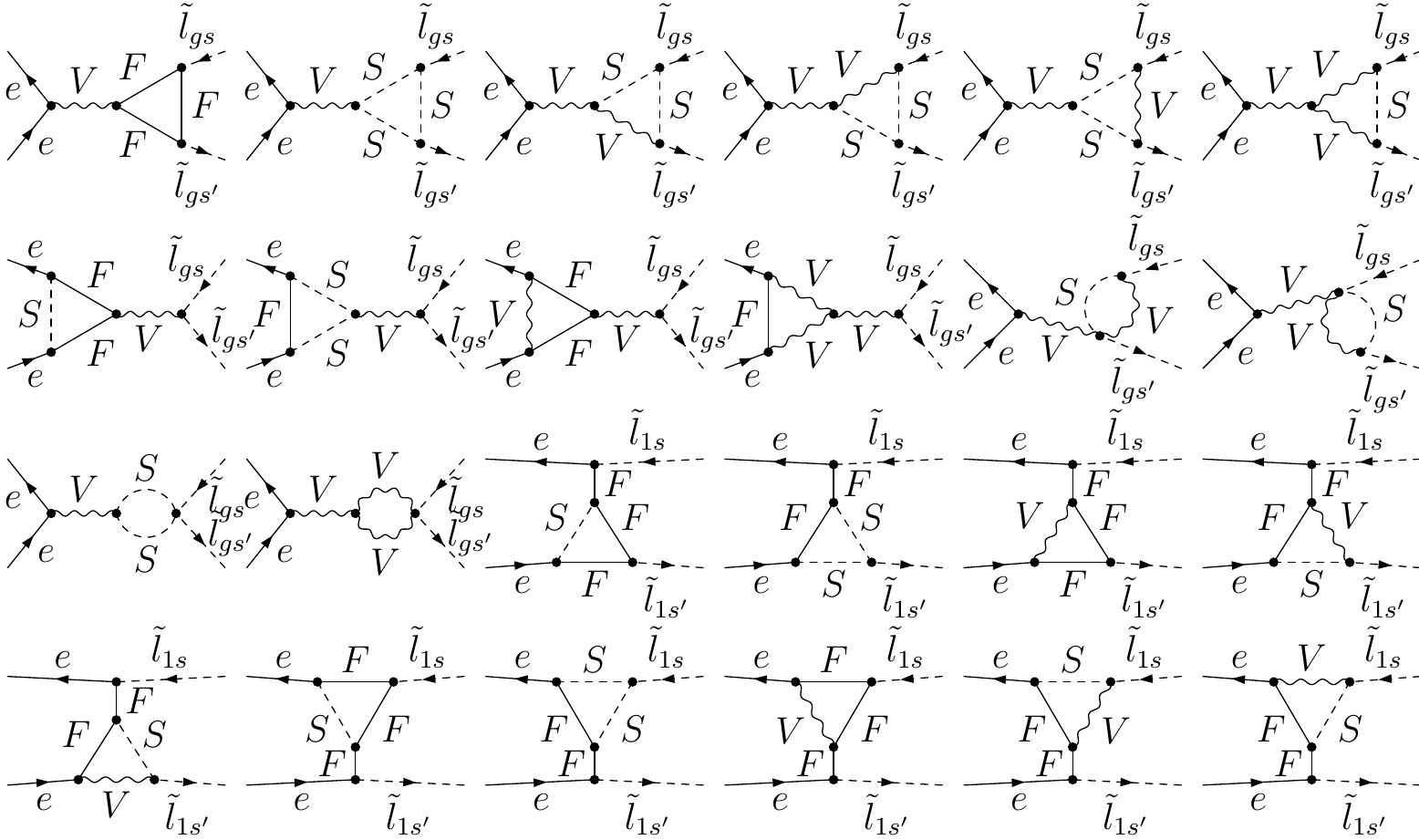}}
\framebox[14cm]{\includegraphics[width=0.75\textwidth]{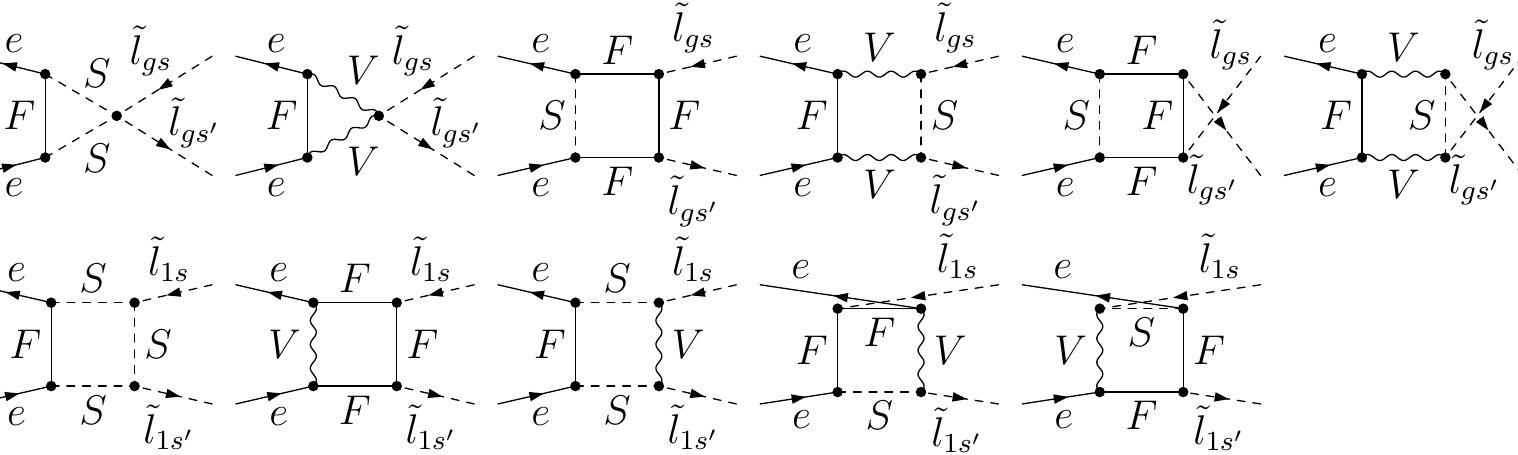}}
\framebox[14cm]{\includegraphics[width=0.75\textwidth]{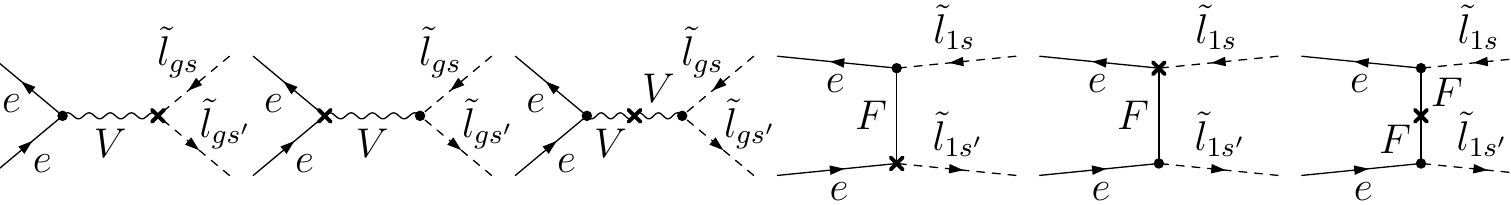}}
\caption{
  Generic tree, self-energy, vertex, box, and counterterm diagrams for the 
  process \eeSlSl\ ($\Sl_{gs} = \{\Se_{gs},\Sn_g\};\; g = 1,2,3;\; s,\spri = 1,2$). 
  The additional diagrams, which occur only in the case of first generation 
  slepton production, are denoted with $\Sl_{1s}$.
  $F$ can be a SM fermion, chargino or neutralino; 
  $S$ can be a sfermion or a Higgs/Goldstone boson; 
  $V$ can be a $\ga$, $Z$ or $W^\pm$. 
  It should be noted that electron--Higgs couplings are neglected. 
}
\label{fig:eeSlSl}
\end{center}
\end{figure}

Sample diagrams for the process \eeSeSe\ and \eeSnSn\ are shown in 
\reffi{fig:eeSlSl}. Diagrams for chargino/neutralino production can be
found in \citere{eeIno}. Not shown in \reffi{fig:eeSlSl} are the
diagrams for real (hard and soft) photon radiation.  
We have neglected all electron--Higgs couplings and terms proportional 
to the electron mass whenever this is safe, \ie\ except when the electron 
mass appears in negative powers or in loop integrals.
We have verified numerically that these contributions are indeed totally 
negligible.  
Moreover, in general, in \reffi{fig:eeSlSl} we have omitted diagrams 
with self-energy type corrections of external (on-shell) particles. 
While the contributions from the real parts of the loop functions are 
taken into account via the renormalization constants defined by OS 
renormalization conditions, the contributions coming from the imaginary 
part of the loop functions can result in an additional (real) correction 
if multiplied by complex parameters.  In the analytical and numerical 
evaluation, these diagrams have been taken into account via the 
prescription described in \citere{MSSMCT}.

As regularization scheme for the UV divergences we have used constrained 
differential renormalization~\cite{cdr}, which has been shown to be 
equivalent to dimensional reduction~\cite{dred1,dred2} at the \onel\ 
level~\cite{formcalc1}. 
Thus the employed regularization scheme preserves SUSY~\cite{dredDS,dredDS2}
and guarantees that the SUSY relations are kept intact.
All UV divergences cancel in the final result.
For a discussion on soft photon emission and corresponding problems with the
phase space integration, see \citeres{eeIno,eeSlep}.


\section{Numerical analysis}
\label{sec:numeval}

Here we review two examples for the numerical analysis of chargino/neutralino
and slepton production at $e^+e^-$ colliders in the cMSSM as presented in
\citeres{eeIno,eeSlep}. 
In the figures below we show the cross sections at the tree level 
(``tree'') and at the full one-loop level (``full''), which is the cross 
section including \textit{all} one-loop corrections.
All results shown use the \Code{CCN[1]} renormalization scheme~\cite{MSSMCT}
(\ie\ OS conditions for the two charginos and the lightest neutralino).

\subsection{The processes \boldmath{\eecc} and \boldmath{\eenn}}
\label{sec:eeccnn}

The SUSY parameters for the evaluation of these production cross sections 
are chosen according to the scenario \Sce, shown in \refta{tab:para-ccnn}.

\begin{table}[htb!]
\centering
\begin{tabular}{lrrrrrrrrrrrr}
\hline
Scen. & $\sqrt{s}$ & $\TB$ & $\mu$ & $\MHp$ & $M_{\tilde Q, \tilde U, \tilde D}$ & 
$M_{\tilde L, \tilde E}$ & $|\At|$ & $\Ab$ & $A_{\tau}$ & 
$|M_1|$ & $M_2$ & $M_3$ \\ 
\hline
\Sce & 1000 & 10 & 450 & 500 & 1500 & 1500 & 2000 & $|\At|$ &
$M_{\tilde L}$ & $\mu$/4 & $\mu$/2 & 2000 \\
\hline
\end{tabular}
\caption{\label{tab:para-ccnn}
  MSSM default parameters for the numerical investigation of chargino
  and neutralino production; all parameters 
  (except of $\TB$) are in GeV.  
}
\end{table}

\begin{figure}[htb!]
\begin{center}
\begin{tabular}{c}
\includegraphics[width=0.48\textwidth,height=5cm]{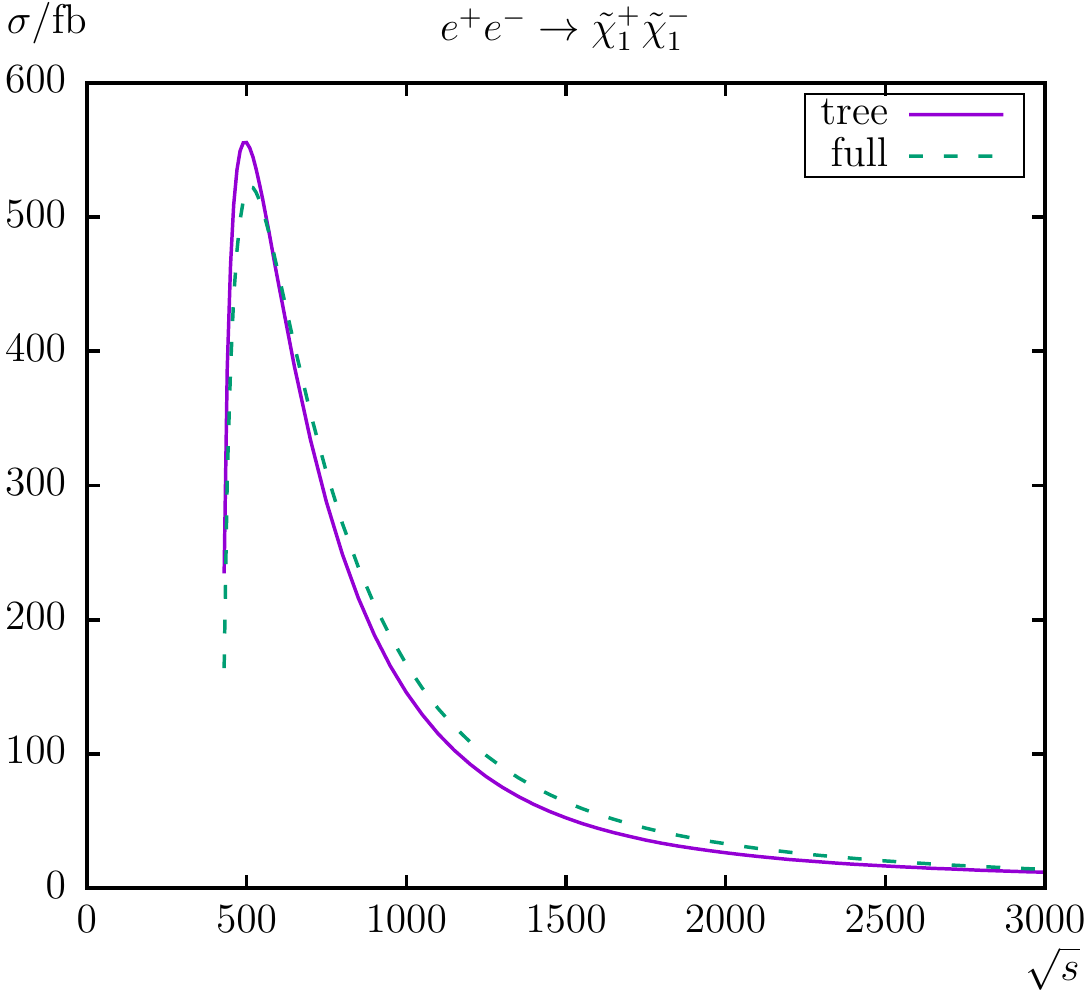}
\includegraphics[width=0.48\textwidth,height=5cm]{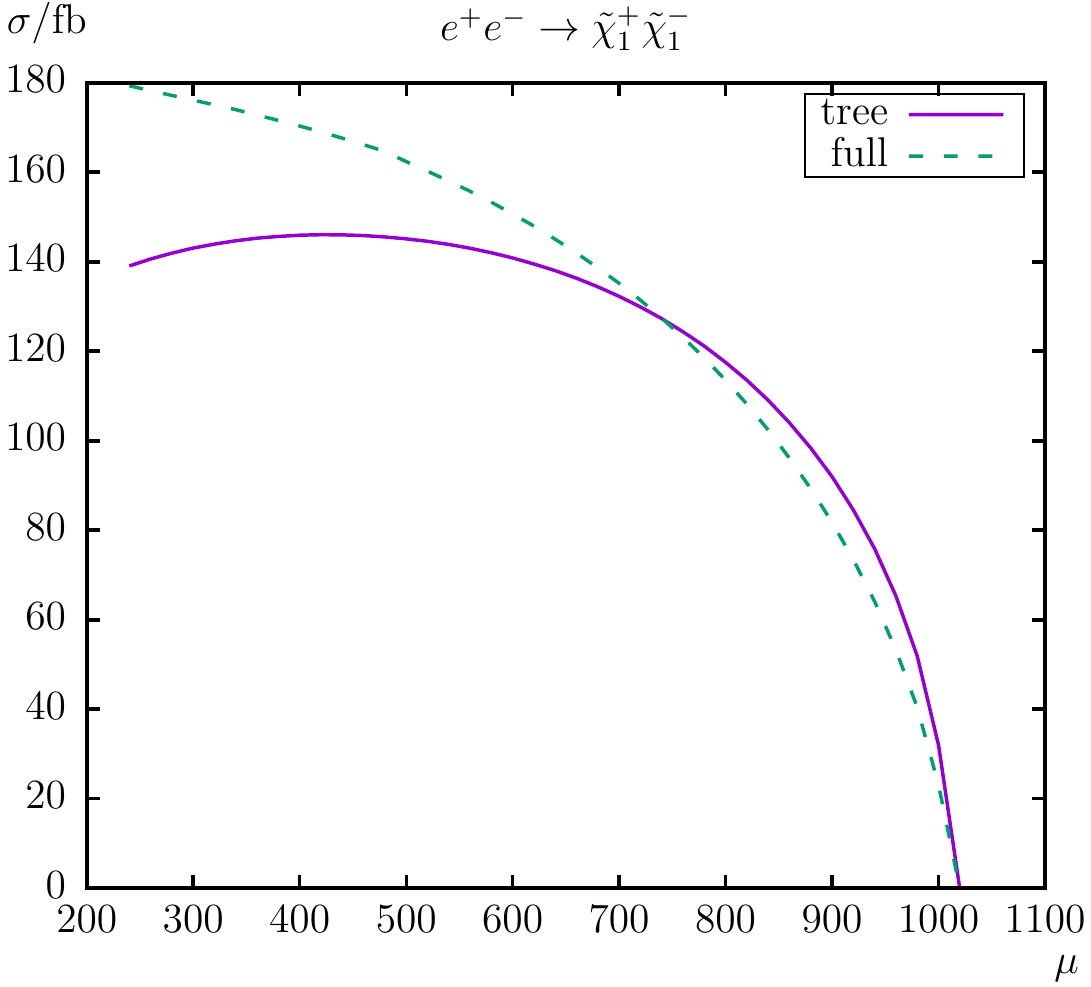}
\\[1em]
\includegraphics[width=0.48\textwidth,height=5cm]{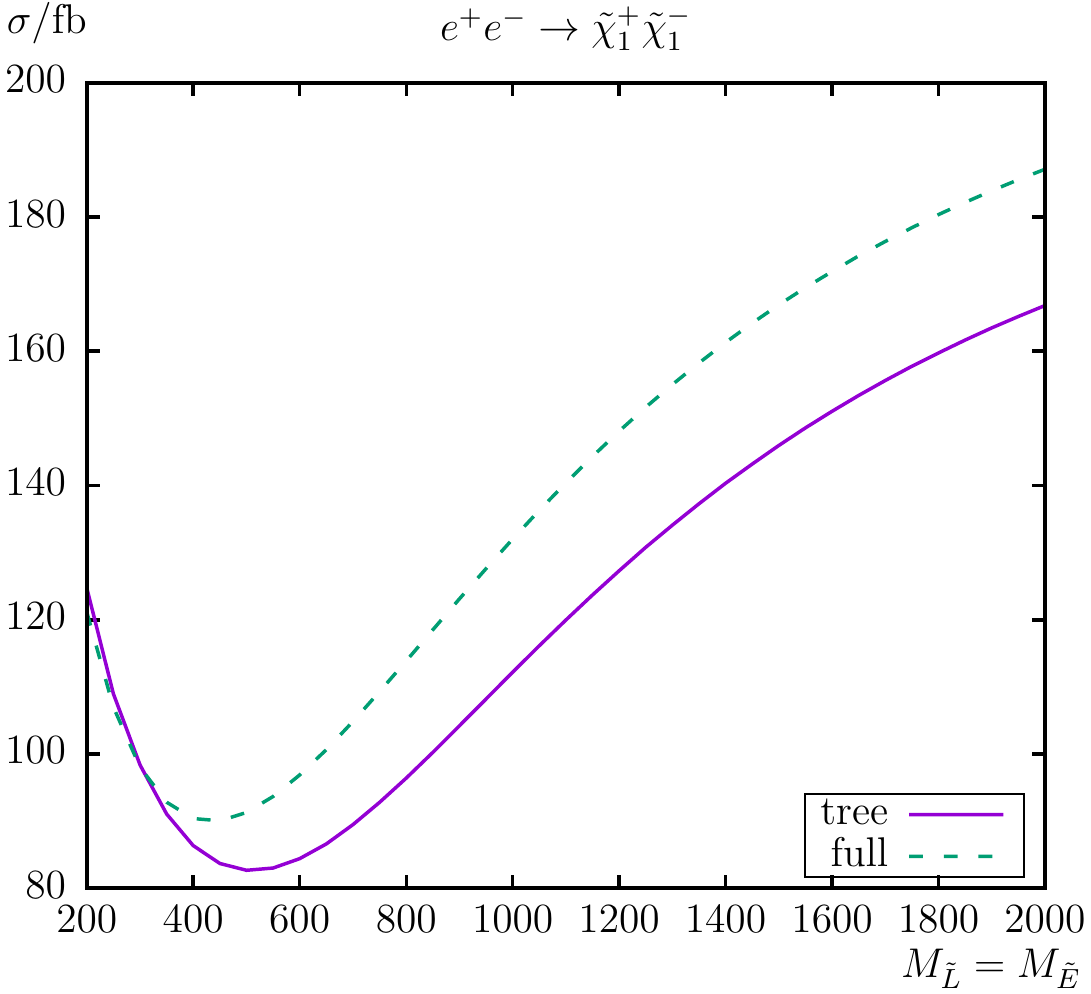}
\includegraphics[width=0.48\textwidth,height=5cm]{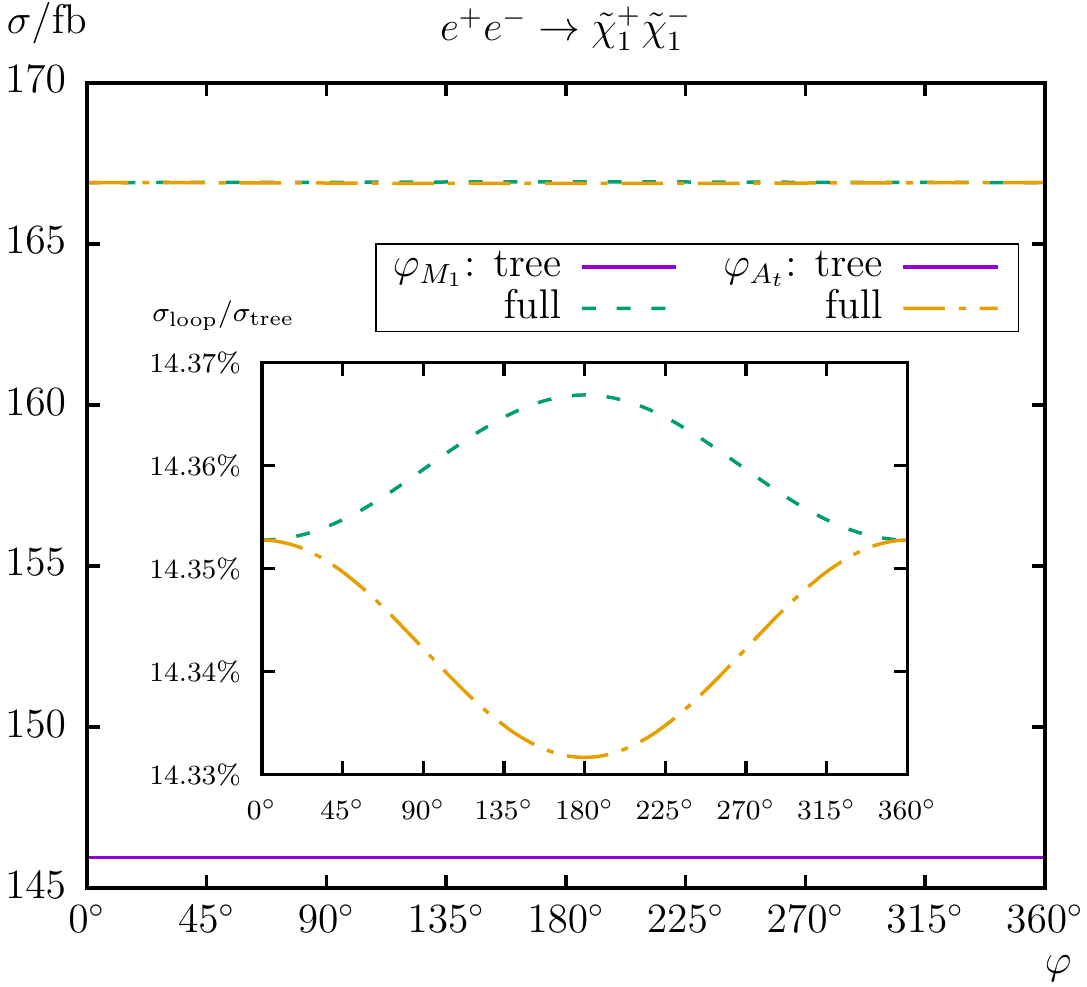}
\end{tabular}
\caption{\label{fig:eec1c1}
  $\sig(\eecece)$.
  Tree-level and full one-loop corrected cross sections are shown with 
  parameters chosen according to \Sce.
  The upper plots show the cross sections with $\sqrt{s}$ (left) and 
  $\mu$ (right) varied;  the lower plots show $\MSL = \MSE$ (left) and 
  $\phiMe$, $\phiAt$ (right) varied.
}
\end{center}
\end{figure}

As an example for chargino/neutralino production the process $\eecece$ is
shown in \reffi{fig:eec1c1}.  
In the analysis of the production cross section as a function of $\sqrt{s}$ 
(upper left plot) we find the expected behavior: a strong rise close to the 
production threshold, followed by a decrease with increasing $\sqrt{s}$. 
Away from the production threshold, loop corrections of $\sim -8\,\%$ at 
$\sqrt{s} = 500\gev$ and $\sim +14\,\%$ at $\sqrt{s} = 1000\gev$ are found 
in scenario \Sce, with a ``tree crossing'' 
(\ie\ where the loop corrections become approximately zero and therefore 
cross the tree-level result) at $\sqrt{s} \approx 575\gev$.
The relative size of loop corrections increase with increasing $\sqrt{s}$ 
(and decreasing $\sig$) and reach $\sim +19\,\%$ at $\sqrt{s} = 3000\gev$.
With increasing $\mu$ in \Sce\ (upper right plot) we find a strong decrease 
of the production cross section, as can be expected from kinematics.
The relative loop corrections in \Sce\ reach $\sim +30\,\%$ 
at $\mu = 240\gev$ (at the border of the experimental limit), $\sim +14\,\%$ 
at $\mu = 450\gev$ (\ie\ \Sce) and $\sim -30\,\%$ at $\mu = 1000\gev$. 
In the latter case these large loop corrections are due to the (relative) 
smallness of the tree-level results, which goes to zero for $\mu = 1020\gev$
(\ie\ the chargino production threshold).
The cross section as a function of $\MSL$ ($= \MSE$) is shown in the lower 
left plot of \reffi{fig:eec1c1}.  This mass parameter controls the
$t$-channel exchange of first generation sleptons at tree-level.
First a small decrease down to $\sim 90$~fb can be observed for 
$\MSL \approx 400\gev$.  For larger $\MSL$ the cross section rises up 
to $\sim 190$~fb for $\MSL = 2\tev$.
In scenario \Sce\ we find a substantial increase of the cross sections from 
the loop corrections.  They reach the maximum of $\sim +18\,\%$ at 
$\MSL \approx 850\gev$ with a nearly constant offset of about $20$~fb
for higher values of $\MSL$.
We find that the phase dependence $\phiMe$ of the cross section in our
scenario is tiny.  The loop corrections are found to be nearly independent 
of $\phiMe$ at the level below $\sim +0.1\,\%$ in \Sce.
We also show the variation with $\phiAt$, which enter via final state 
vertex corrections.  While the variation with $\phiAt$ is somewhat larger 
than with $\phiMe$, it remains tiny and unobservable.
However, in \citere{eeIno} other production channels with an appreciable
phase dependence were identified.


\subsection{The processes \boldmath{\eeSeSe} and \boldmath{\eeSnSn}}
\label{sec:eeslepslep}

The SUSY parameters for the numerical analysis for slepton production
(\ie\ in \citere{eeSlep}) are chosen according to the scenario \Scz, shown 
in \refta{tab:para-slep}.

\begin{table}[htb!]
\centering
\begin{tabular}{lrrrrrrrrrrrr}
\hline
Scen. & $\sqrt{s}$ & $\TB$ & $\mu$ & $\MHp$ & $M_{\tilde Q, \tilde U, \tilde D}$ & 
$\MSE$ & $A_{\Fu_g}$ & $A_{\Fd_g}$ & $|A_{\Fe_g}|$ & $|M_1|$ & $M_2$ & $M_3$ \\ 
\hline
\Scz & 1000 & 10 & 350 & 1200 & 2000 & 300 & 2600 & 2000 & 2000 & 400 & 
600 & 2000 \\
\hline
\end{tabular}
\caption{\label{tab:para-slep}
  MSSM default parameters for the numerical investigation; all 
  parameters (except of $\TB$) are in GeV. Furthermore, 
$\MSL = \MSE + 50\gev$ was chosen for all slepton generations.  
}
\end{table}

\begin{figure}[htb!]
\begin{center}
\begin{tabular}{c}
\includegraphics[width=0.48\textwidth,height=5cm]{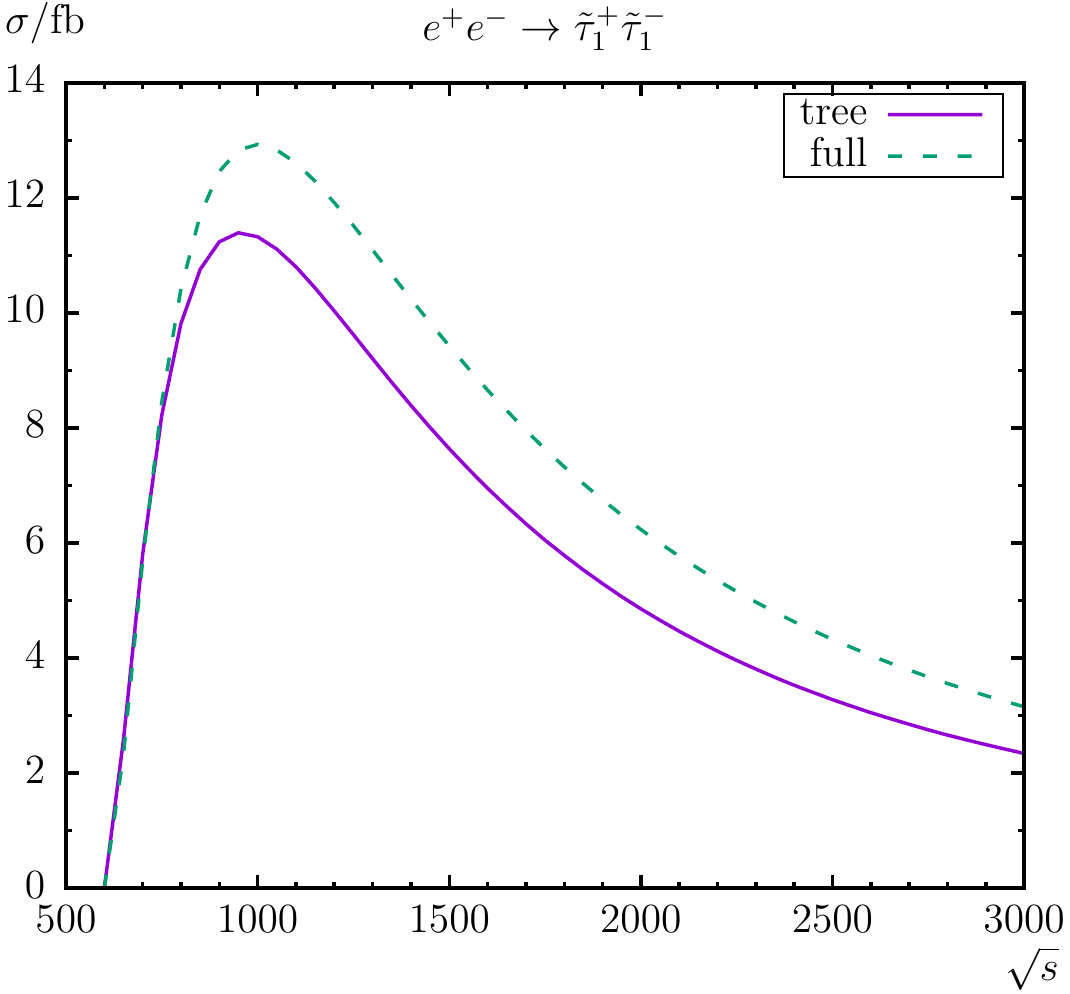}
\includegraphics[width=0.48\textwidth,height=5cm]{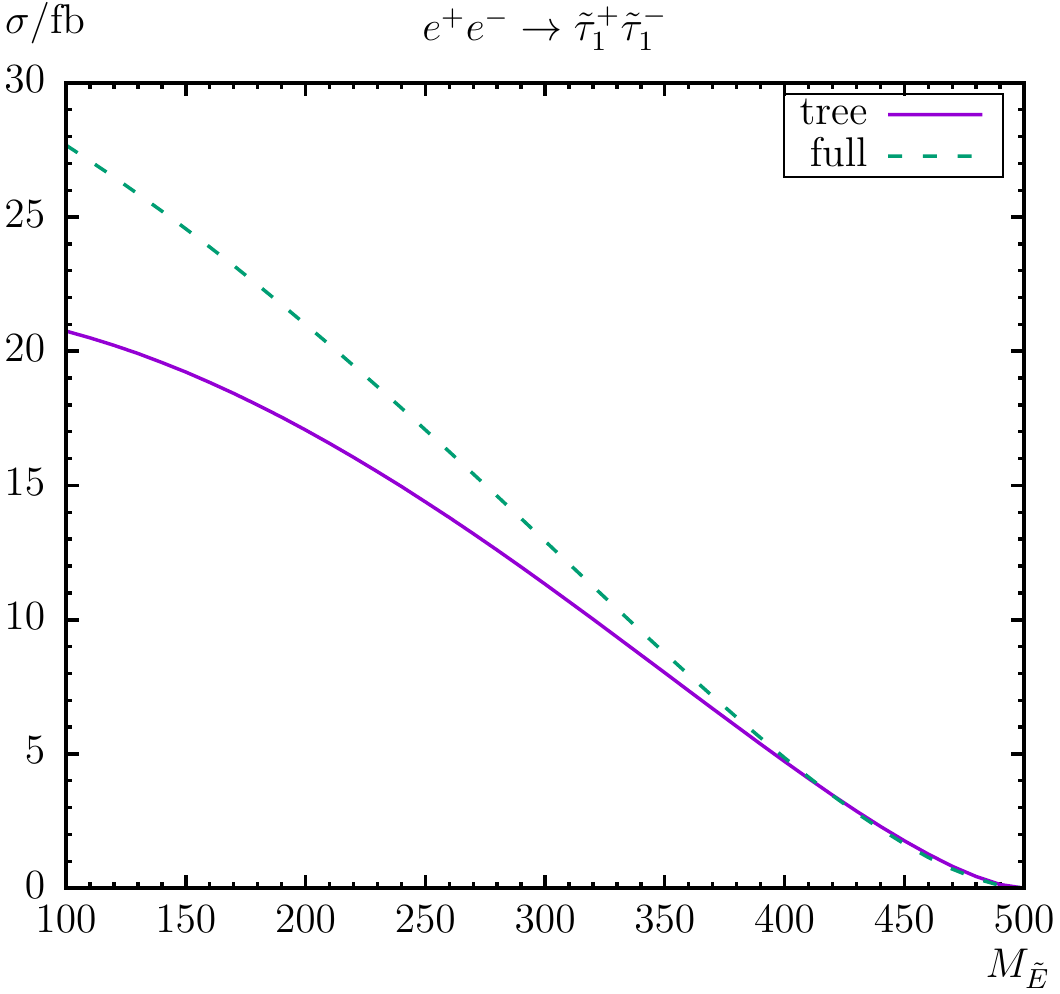}
\\[1em]
\includegraphics[width=0.48\textwidth,height=5cm]{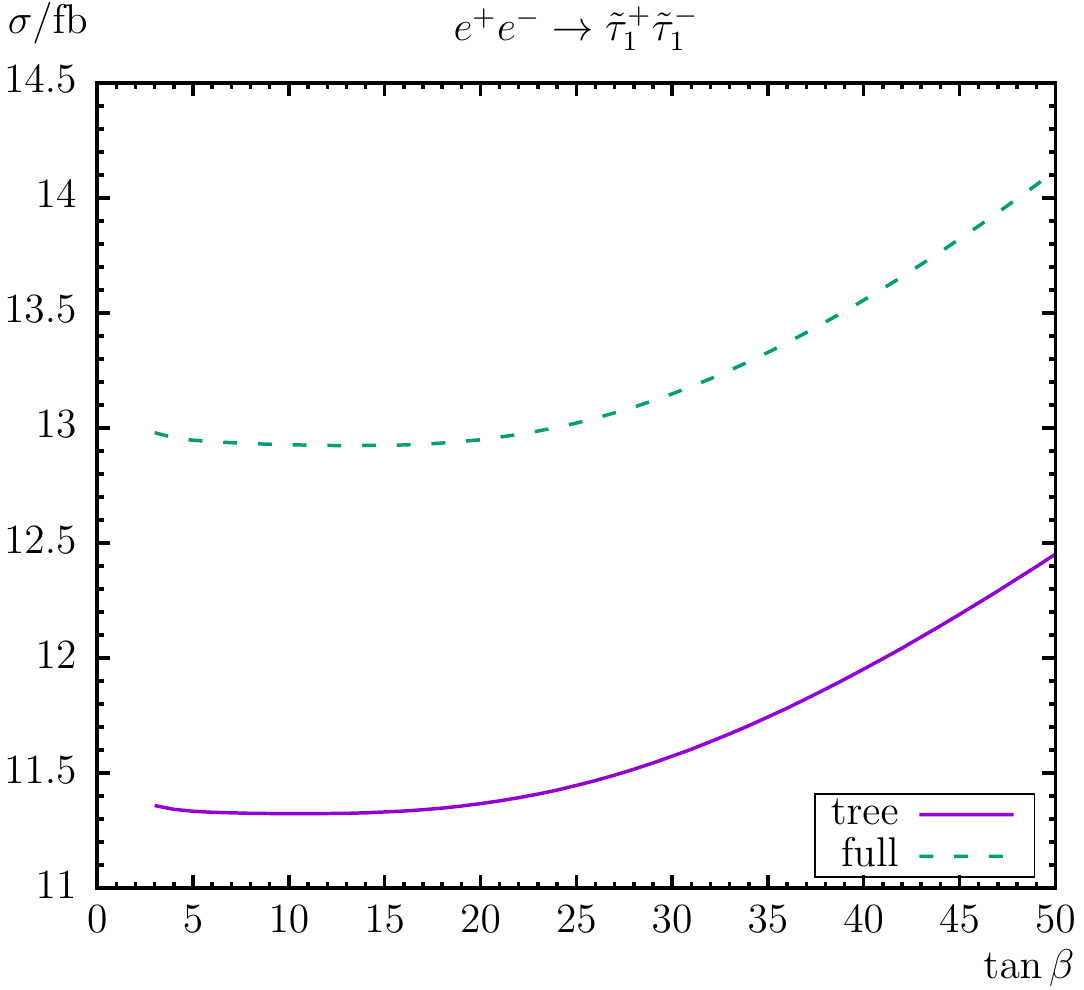}
\includegraphics[width=0.48\textwidth,height=5cm]{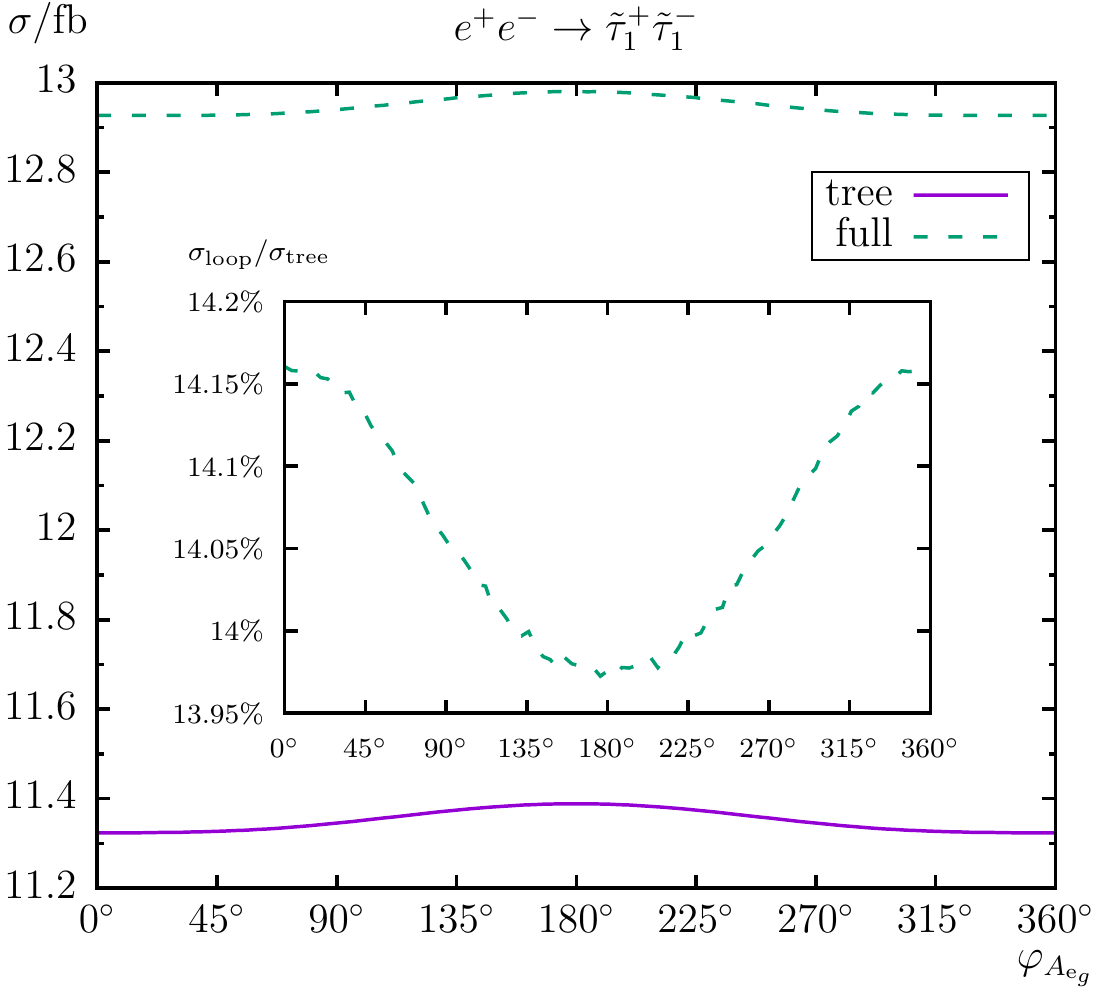}
\\
\end{tabular}
\caption{\label{fig:eeSa1Sa1}
  $\sig(\eeSaeSae)$.
  Tree-level and full one-loop corrected cross sections are shown 
  with parameters chosen according to \Scz.
  The upper plots show the cross sections with $\sqrt{s}$ (left) 
  and $\MSE$ (right) varied; the lower plots show $\TB$ (left) and 
  $\phiAeg$ (right) varied. All masses and energies are in GeV.
}
\end{center}
\end{figure}

As an example of the numerical analysis presented in 
\citere{eeSlep} we show the 
process $\eeSaeSae$ in \reffi{fig:eeSa1Sa1}. 
As a function of $\sqrt{s}$ we find loop corrections of $\sim +14\,\%$ at 
$\sqrt{s} = 1000\gev$ (\ie\ \Scz), a tree crossing at 
$\sqrt{s} \approx 725\gev$ (where the one-loop corrections are
between $\pm 10\,\%$ for $\sqrt{s} \lsim 900 \gev$) and $\sim +35\,\%$ 
at $\sqrt{s} = 3000\gev$.
In the analysis as a function of $\MSE$ (upper right plot) the cross 
sections are decreasing with increasing $\MSE$ as obvious from 
kinematics and the full corrections have their maximum of $\sim 28\,\fb$ 
at $\MSE = 100\gev$, more than two times larger than in \Scz.  The relative 
corrections are changing from $\sim +33\,\%$ at $\MSE = 100\gev$ to 
$\sim -25\,\%$ at $\MSE = 490\gev$ with a tree crossing at $\MSE = 415\gev$.
In the lower left row of \reffi{fig:eeSa1Sa1} we show the dependence 
on $\TB$. The relative corrections for the $\TB$ dependence vary between 
$\sim +14.2\,\%$ at $\TB = 5$ and $\sim +13.4\,\%$ at $\TB = 50$.

The phase dependence $\phiAeg$ of the cross section in \Scz\ is shown 
in the lower right plot of \reffi{fig:eeSa1Sa1}.  
The loop correction increases the tree-level result by $\sim +14\,\%$.
The phase dependence of the relative loop correction is very small and 
found to be below $0.2\,\%$.
The variation with $\phiMe$ is negligible and therefore not shown here.


\subsection*{Acknowledgements}

S.H.\ thanks the  organizers of L\&L\,2018 for the  invitation and the
(as always!) inspiring atmosphere.
The work of S.H.\ is supported 
in part by the MEINCOP Spain under contract FPA2016-78022-P, 
in part by the ``Spanish Agencia Estatal de Investigaci\'on'' (AEI) and the EU
``Fondo Europeo de Desarrollo Regional'' (FEDER) through the project
FPA2016-78022-P, 
in part by the ``Spanish Red Consolider MultiDark'' FPA2017‐90566‐REDC, 
and in part by the AEI through the grant IFT
Centro de Excelencia Severo Ochoa SEV-2016-0597.


\newcommand\jnl[1]{\textit{\frenchspacing #1}}
\newcommand\vol[1]{\textbf{#1}}

\end{document}